\documentclass[preprint,review,3p,12pt]{elsarticle}

\usepackage{amsfonts,amsmath,amssymb,mathrsfs,bbm,bm}
\usepackage{graphics}
\biboptions{comma,sort&compress}

\journal{Physica A}

\begin{document}

\begin{frontmatter}

\title{Complexity-entropy causality plane: a useful approach for distinguishing songs}

\author[UEM,NU]{Haroldo V. Ribeiro}
\ead{hvr@dfi.uem.br}
\author[CONICET,UNLP]{Luciano Zunino}
\author[UEM]{Renio S. Mendes}
\author[UEM]{Ervin K. Lenzi}
\address[UEM]{Departamento de F\'isica and National Institute of Science and Technology for Complex Systems, Universidade Estadual de Maring\'a, Av. Colombo 5790, 87020-900, Maring\'a, PR, Brazil}
\address[NU]{Department of Chemical and Biological Engineering, Northwestern University, Evanston, IL 60208, USA}
\address[CONICET]{Centro de Investigaciones \'Opticas (CONICET La Plata - CIC), C.C. 3, 1897 Gonnet, Argentina}
\address[UNLP]{Departamento de Ciencias B\'asicas, Facultad de Ingenier\'ia, Universidad Nacional de La Plata (UNLP), 1900 La Plata, Argentina}

\begin{abstract}
Nowadays we are often faced with huge databases resulting from the rapid growth of data storage technologies. This is particularly true when dealing with music databases. In this context, it is essential to have techniques and tools able to discriminate properties from these massive sets. In this work, we report on a statistical analysis of more than ten thousand songs aiming to obtain a complexity hierarchy. Our approach is based on the estimation of the permutation entropy combined with an intensive complexity measure, building up the complexity-entropy causality plane. The results obtained indicate that this representation space is very promising to discriminate songs as well as to allow a relative quantitative comparison among songs. Additionally, we believe that the here-reported method may be applied in practical situations since it is simple, robust and has a fast numerical implementation.
\end{abstract}


\end{frontmatter}

\section{Introduction}
Nowadays we are experimenting a rapid development of technologies related to data storage. As an immediate consequence, we are often faced with huge databases hindering the access to information. Thus, it is necessary to have techniques and tools able to discriminate elements from these massive databases. Text categorization~\cite{Sebastiani}, scene classification~\cite{Radke} and protein classification~\cite{Enright} are just a few examples where this problem emerges. In a parallel direction, statistical physicists are increasingly interested in studying the so-called complex systems~\cite{Auyang,Jensen,Barabasi,Sornette,Boccara}. These investigations employ established methods of statistical 
mechanics as well as recent developments of this field aiming to extract hidden patterns that are governing the system's dynamics. In a similar way, this framework may help to advance in distinguishing elements within these databases, with the benefit of the simplicity often attributed to statistical physics methods.

A very interesting case corresponds to the music databases, not only because of the incredible amount of data (for instance, the iTunes Store has more than 14 million songs), but also due to the ubiquity of music in our society as well as its deeply connection with cognitive habits and historical developments~\cite{DeNora}. In this direction, there are investigations focused on collective listening habits~\cite{Lambiotte,Lambiotte2,Buldu}, collaboration networks among artists~\cite{Teitelbaum}, music sales~\cite{Lambiotte3}, success of musicians~\cite{Davies,Borges,Hu}, among others. On the other hand, the sounds that compose the songs present several complex structures and emergent features which, in some cases,
resemble very closely the patterns of out-of-equilibrium physics, such as scale-free statistics and universality. For instance, the seminal work of Voss and Clarke~\cite{Voss} showed that the power spectrum associated to the loudness variations and pitch fluctuations of radio stations (including songs and human voice) is characterized by $1/f$ noise-like pattern in the low frequency domain ($f \leq 10 Hz$). Klimontovich and Boon~\cite{Klimontovich} argue that this behavior for low-frequency follows from a natural flicker noise theory. However, this finding has been questioned by Nettheim~\cite{Nettheim} and according to him the power spectrum may be better described by $1/f^2$. Fractal structures were also reported by Hs\"u and Hs\"u~\cite{Hsu2,Hsu} when studying classical pieces concerning frequency intervals. It was also found that the distribution of sound amplitudes may be adjusted by a one-parameter stretched Gaussian and that this non-Gaussian feature is related to correlation aspects present in the songs~\cite{Mendes}.

These features and others have attracted the attention of statistical physicists, who have attempted to obtain some quantifiers able to distinguish songs and genres. One of these efforts was made by Jennings et al.~\cite{Jennings} who found that the Hurst exponent estimated from the volatility of the sound intensity depends on the music genre. Correa et al.~\cite{Correa} investigated four music genres employing a complex network representation for rhythmic features of the songs. There are still other investigations~\cite{Boon,Bigerelle,Diodati,Gunduz,Su2,Scaringella,Jafari,Dagdug,Su,Rio,Ro,Serra,Mostafa,Boon2}, most of which are based on fractal dimensions, entropies, power spectrum analysis or correlation analysis. It is worth noting that there are several methods of automatic genre classification emerging from engineering disciplines (see, for instance, Ref.~\cite{Tzanetakis}). {In particular, there exists a very active community working on music classification problems and several important results are published at the ISMIR~\cite{ISMIR} conferences (just to mention a few please see Refs.~\cite{ISMIR1,ISMIR2,ISMIR3,ISMIR4,ISMIR5,ISMIR6,ISMIR7,ISMIR8,ISMIR9,ISMIR10,ISMIR11,ISMIR12,ISMIR13}.)}

However, the music genre it not a well defined concept~\cite{Scaringella}, and, specially, the boundaries between genres still remain fuzzy. Thus, any taxonomy may be controversial, representing a challenging and open problem of pattern recognition. In addition, some of the proposed quantifiers require specific algorithms or recipes for processing the sound of the songs, which may depend on tuning parameters.

Here, we follow an Information Theory approach trying to quantify aspects of songs. More specifically, the Bandt and Pompe approach~\cite{Bandt} is applied in order to obtain a complexity hierarchy for songs. This method defines a ``natural'' complexity measure for time series based on ordinal patterns. Although this concept has not been explored yet within the context of music, it has been successfully applied in other areas, such as medical~\cite{Li,Nicolaou}, financial~\cite{Zunino,Zunino3} and climatological time series~\cite{Saco,Barreiro}. In this direction, our main goal is to fill this hiatus employing the Bandt and Pompe approach together with a non-trivial entropic measure~\cite{LopezRuiz,Martin,Lamberti}, constructing the so-called complexity-entropy causality plane~\cite{Zunino,Zunino3,Rosso,Zunino2}. As it will be discussed in detail below, we have found that this representation space is very promising to distinguish songs from huge databases. Moreover, thanks to the simple and fast implementation it is possible to conjecture its use in practical situations. In the following, we review some aspects related to the Bandt and Pompe approach as well as the complexity-entropy causality plane (Section 2). Next, we describe our database and the results (Section 3). Finally, we end this work with some concluding comments (Section 4).

\section{Methods}
The essence of the permutation entropy proposed by Bandt and Pompe~\cite{Bandt} is to associate a symbolic sequence to the time series under analysis. This is done by employing a suitable partition based on ordinal patterns obtained by comparing neighboring values of the original series. To be more specific, consider a given time series $\{x_t\}_{t=1,\dots,N}$ and the following partitions represented by a $d$-dimensional vector ($d>1, D \in \mathbb{N}$)
\[
(s)\mapsto (x_{s-(d-1)},x_{s-(d-2)},\dots,x_{s-1},x_{s})\;,
\]
with $s=d,d+1,\dots,N$. For each one of these $(N-d+1)$ vectors, we investigate the permutations \mbox{$\pi=(r_0,r_1,\dots,r_{d-1})$} of $(0,1,\dots,d-1)$ defined by $x_{s-r_{d-1}}\leq x_{s-r_{d-2}}\leq \dots \leq x_{s-r_{1}} \leq x_{s-r_{0}}$, and, for all $d\, !$ possible permutations of $\pi$, we evaluate the probability distribution $P=\{p(\pi)\}$ given by
\[
p(\pi) = \frac{\#\{s|s\leq N-d+1;~ (s) ~\text{has type}~ \pi \}}{N-d+1}\;,
\]
where the symbol $\#$ stands for the number (frequency) of occurrences of the permutation $\pi$. Thus, we define the normalized permutation entropy of order by
\begin{equation}
H_s[P]=\frac{S[P]}{\log d\,!}\;,
\end{equation}
with $S[P]$ being the standard Shannon's entropy~\cite{Shannon}. Naturally, $0 \leq H_s[P] \leq 1$, where the upper bound occurs for a completely random system, i.e., a system for which all $d\,!$ possible permutations are equiprobable. If the time series exhibits some kind of ordering dynamics $H_s[P]$ will be smaller than one. As pointed out by Bandt and Pompe~\cite{Bandt}, the advantages in using this method lie on its simplicity, robustness and very fast computational evaluation. Clearly, the parameter $d$ (known as embedding dimension) plays an important role in the estimation of the permutation probability distribution $P$, since it determines the number of accessible states. In fact, the choice of $d$ depends on the length $N$ of the time series in such a way that the condition $d\,!\ll N$ must be satisfied in order to obtain a reliable statistics. For practical purposes, Bandt and Pompe recommend $d=3,\dots,7$. Here, we have fixed $d=5$ because the time series under analysis are large enough (they have more than one million of data values). {We have verified that the results are robust concerning the choice of the embedding dimension $d$.} 

Advancing with this brief revision, we now introduce another statistical complexity measure able to quantify the degree of physical structure present in a time series~\cite{LopezRuiz,Martin,Lamberti}. Given a probability distribution $P$, this quantifier is defined by the product of the normalized entropy $H_s$, and a suitable metric distance between $P$ and the uniform distribution $P_e=\{1/d\,!\}$. Mathematically, we may write
\begin{equation}
C_{js}[P]=Q_j[P,P_e]\,H_s[P]\,,
\end{equation}
where
\begin{equation}
Q_j[P,P_e] = \frac{S[(P+P_e)/2] - S[P]/2 - S[P_e]/2}{Q_{\text{max}}}\,
\end{equation}
and $Q_{\text{max}}$ is the maximum possible value of $Q_j[P,P_e]$, obtained when one of the components of $P$ is equal to one and all the others vanish, i.e.,
\[
Q_{\text{max}}=-\frac{1}{2}\left[ \frac{d\,!+1}{d\,!} \log(d\,!+1) - 2 \log(2 d\,!) + \log(d\,!) \right]\,.
\]
The quantity $Q_j$, usually known as disequilibrium, will be different from zero if there are more likely states among the accessible ones. It is worth noting that the complexity measure $C_{js}$ is not a trivial function of the entropy~\cite{LopezRuiz} because it depends on two different probability distributions, the one associated to the system under analysis, $P$, and the uniform distribution, $P_e$. It quantifies the existence of correlational structures, providing important additional information that may not be carried only by the permutation entropy. Furthermore, it was shown that for a given $H_s$ value, there exists a range of possible $C_{js}$ values~\cite{Martin2}.
Motivated by the previous discussion, Rosso et al.~\cite{Rosso} proposed to employ a diagram of $C_{js}$ versus $H_s$ for distinguishing between stochasticity and chaoticity. This representation space, called complexity-entropy causality plane~\cite{Rosso,Zunino,Zunino3}, herein will be our approach for distinguishing songs.

{The concept of ordinal patterns can be straightforward generalized for non-consecutive samples, introducing a lag of $\tau$ (usually known as embedding delay) sampling times. With $\tau=1$ the consecutive case is recovered, and the analysis focuses on the highest frequency contained within the time series. It is clear that different time scales are taken into account by changing the embedding delays of the symbolic reconstruction. The importance of selecting an appropriate embedding delay in the estimation of the permutation quantifiers has been recently confirmed for different purposes, like identifying intrinsic time scales of delayed systems~\cite{zunino2010,soriano2011}, quantifying the degree of unpredictability of the high-dimensional chaotic fluctuations of a semiconductor laser subject to optical feedback~\cite{zunino2011}, and classifying cardiac biosignals~\cite{parlitz2011}. We have found that an embedding delay $\tau=1$ is the optimal one for our music categorization goal since when this parameter is increased the permutation entropy increases and the permutation statistical complexity decreases. Thus, the range of variation of both quantifiers is smaller and, consequently, it is more difficult to distinguish songs and genres.}

\section{Data Presentation and Results}
It is clear that a music piece can be naturally considered as the time evolution of an acoustic signal and time irreversibility is inherent to musical expression~\cite{Boon,Boon2}. From the physical point of view, the songs may be considered as pressure fluctuations traveling through the air. These waves are perceived by the auditory system leading the sense of hearing. In the case of recordings, these fluctuations are converted into a voltage signal by a record system
and then stored, for instance, in a compact disc (CD). The perception of sound is usually limited to a certain range of frequencies - for human beings the full audible range is approximately between 20 Hz and 20 kHz. Because of this limitation the record systems often employ a sampling rate of 44.1 kHz encompassing all the previous spectrum. All the songs analyzed here have this sampling rate.

Our database consists of 10124 songs distributed into ten different music genres, they are: blues (1020), classical (997), flamenco (679), hiphop (1000), jazz (700), metal (1638), Brazilian popular music - mpb (580), pop (1000), tango (1016) and techno (1494). The songs were chosen aiming to cover a large number of composers and singers. To achieve this {and also to determine the music genre via an external judgment}, we tried to select CDs that are compilations of a given genre 
or from representative musical groups {of a given genre}. 


\begin{figure}[!t]
\centering
\includegraphics[scale=0.67]{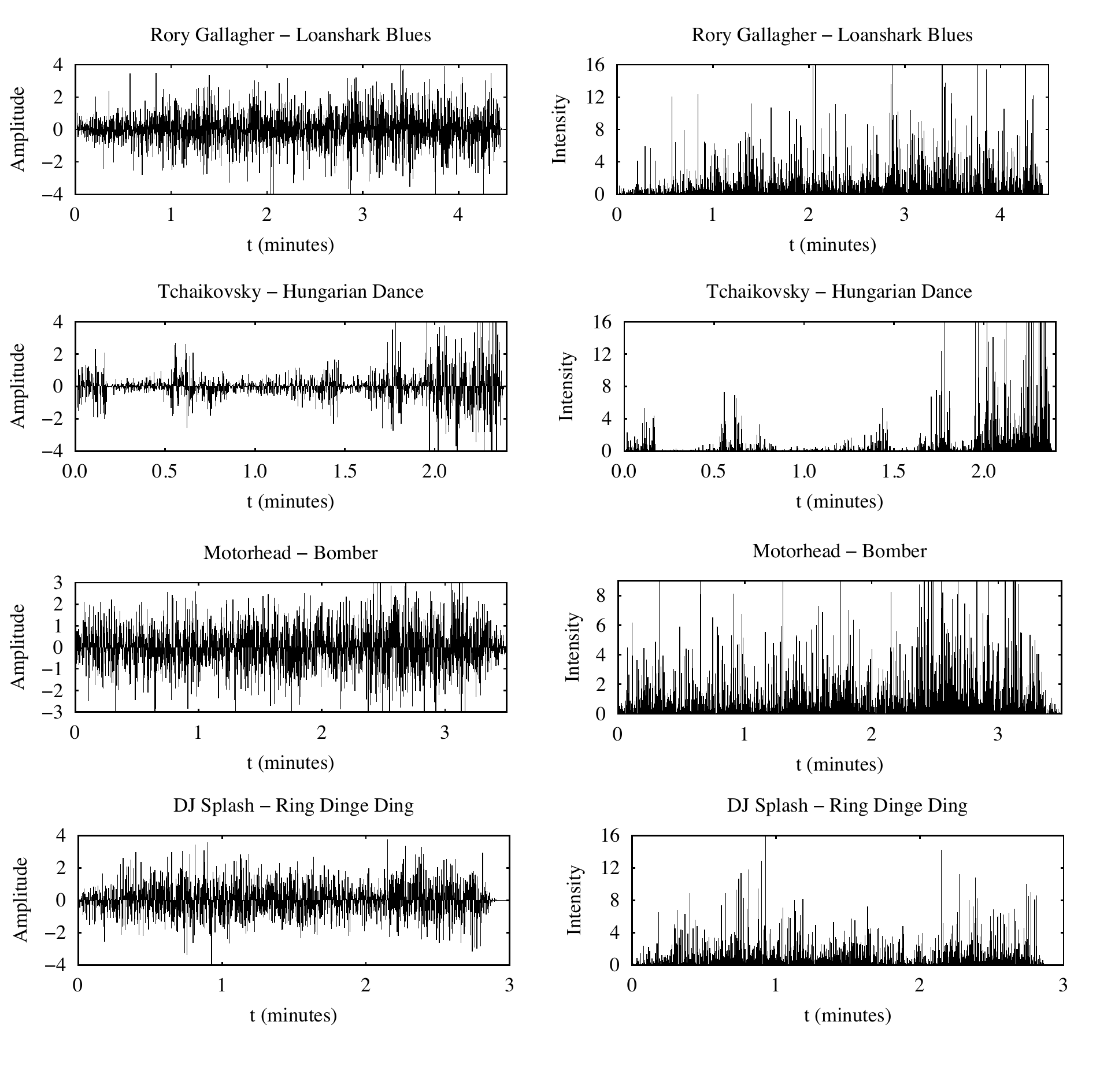}
\caption{A graphical representation of 4 songs from 4 different genres. In the left panel we show the amplitude series and in the right panel the intensity series. The music genres are blues, classic, metal and techno, respectively.}\label{fig:sample}
\end{figure}

\begin{figure}[!t]
\centering
\includegraphics[scale=0.78]{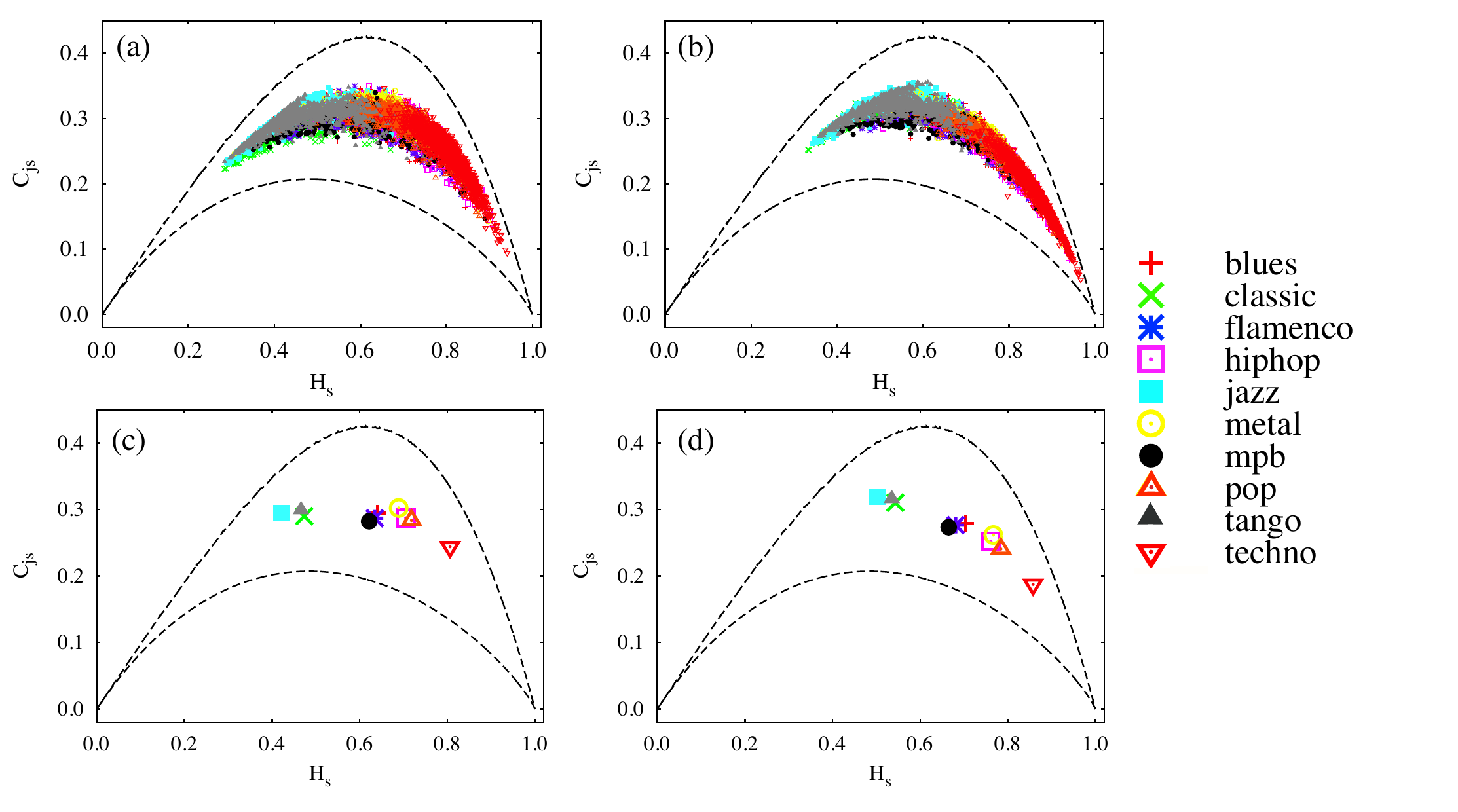}
\caption{(color online) Complexity-entropy causality plane, i.e., $C_{js}$ versus $H_s$ for all the songs when considering the (a) amplitude series and (b) the intensity series. In (c) and (d), we show the mean value of $C_{js}$ and $H_s$ for each genre. The upper (bottom) dashed line represents the maximum (minimum) value of $C_{js}$ as a function of $H_s$ for $d=5$ and the different symbols refer to the 10 different genres. For a better visualization of the different genres see also Figs. \ref{fig:ampgenre} and \ref{fig:intgenre}.}\label{fig:plane}
\end{figure}

By using the previous database, we focus our analysis on two times series directly obtained from the digitized files that represent each song - the sound amplitude series and the sound intensity series, i.e., the square of the amplitude. Figure~\ref{fig:sample} shows these two time series for several songs. We evaluate the normalized entropy $H_s$ and the statistical complexity measure $C_{js}$ for the amplitude and intensity series associated to each song as shown in Figs.~\ref{fig:plane}a and \ref{fig:plane}b. Notice that both series, amplitude and intensity, lead to similar behavior, contrarily to what happens with other quantifiers. For instance, when dealing with Hurst exponent is preferable to work with the intensities~\cite{Mendes} or volatilities~\cite{Jennings}, since the amplitudes are intrinsically anti-correlated due the oscillatory nature of the sound. Moreover, we have found that there is a large range of $H_s$ and $C_{js}$ possible values. This wide variation allows a relative comparison among songs and someone may ask to listen songs that are limited within some interval of $H_s$ and/or $C_{js}$ values. We also evaluate the mean values of $C_{js}$ and $H_s$ over all songs grouped by genre as shown by Figs. \ref{fig:plane}c and \ref{fig:plane}d. These mean values enable us to quantify the complexity of each music genre. In particular, we can observe that high art music genres (e.g. classic, jazz and tango) are located in the central part of the complexity plane, being equally distant from the fully aleatory limit ($H_s\to1$ and $C_{js}\to0$) and also from the completely regular case ($H_s\to0$ and $C_{js}\to0$). On the other hand, light/dance music genres (e.g. pop and techno) are located closer to the fully aleatory limit (white noise). In this context, our approach agrees with other works~\cite{Mendes,Jennings,Diodati}.

\begin{figure}[!t]
\centering
\includegraphics[scale=0.45]{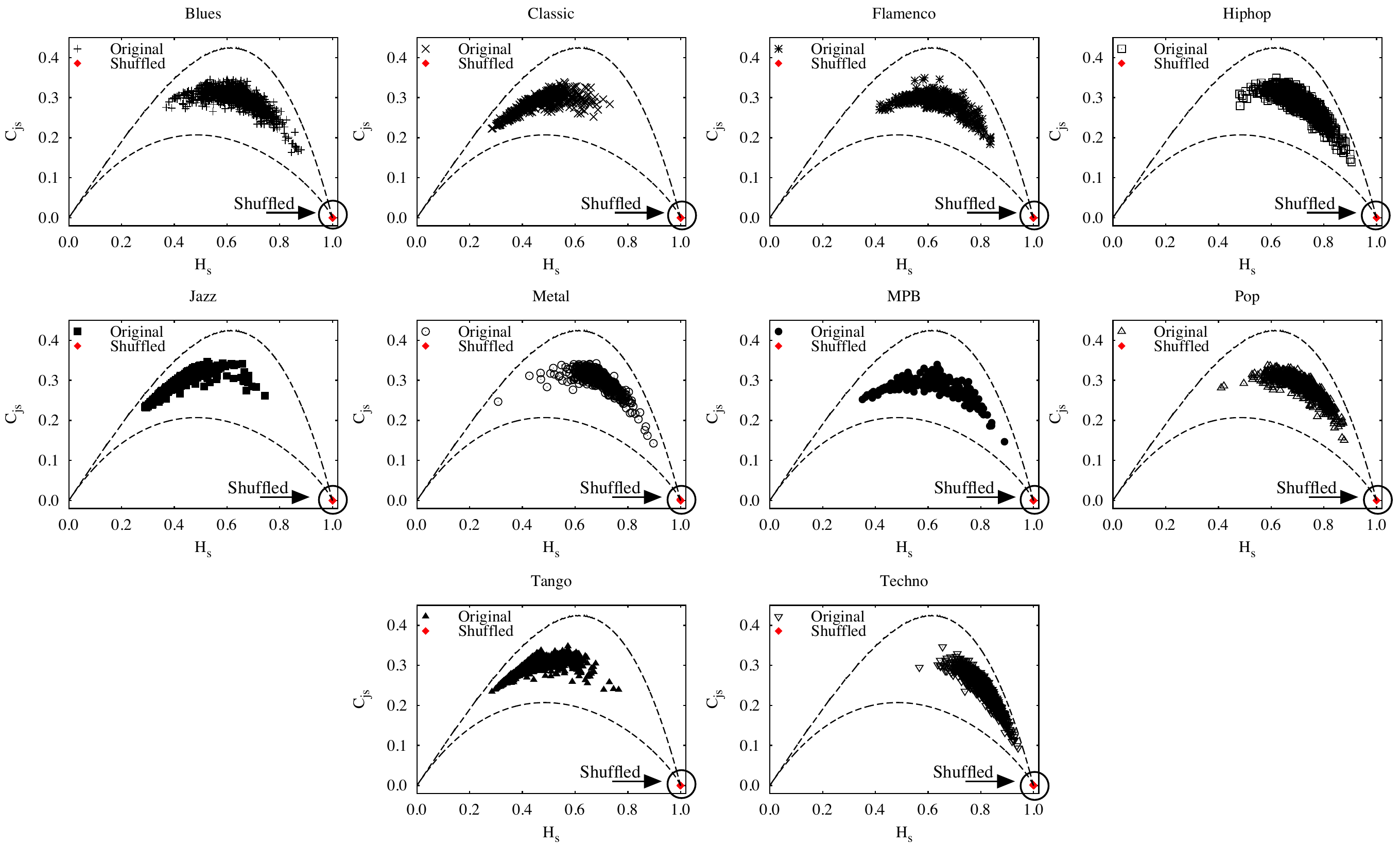}
\caption{Complexity-entropy causality plane for the amplitude series by music genres when considering the original and shuffled series. The upper (bottom) dashed line represents the maximum (minimum) value of $C_{js}$ as a function of $H_s$ for $d=5$ {and the arrows are indicating the shuffled analysis}.}\label{fig:ampgenre}
\end{figure}

\begin{figure}[!ht]
\centering
\includegraphics[scale=0.45]{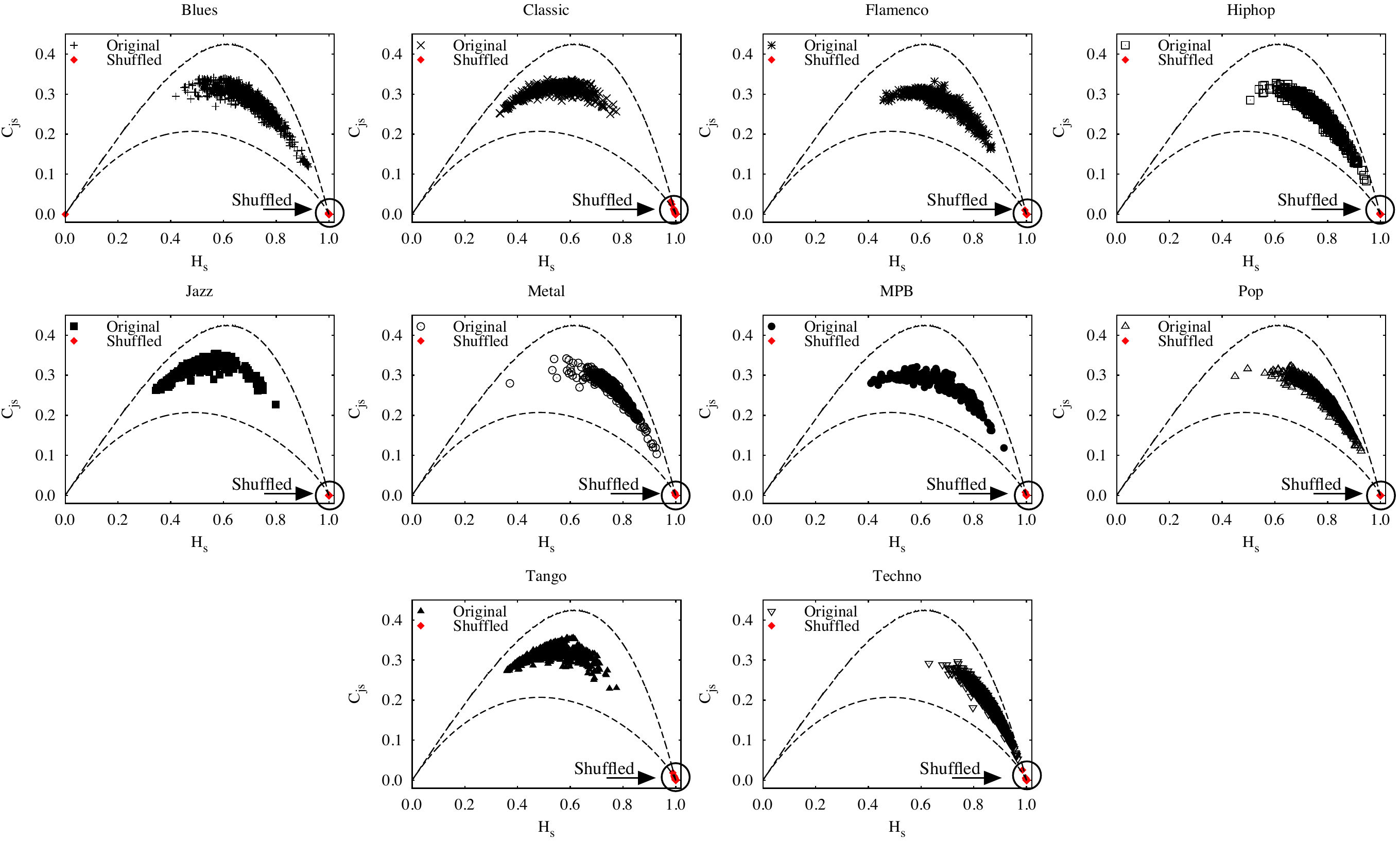}
\caption{Complexity-entropy causality plane for the intensity series by music genres when considering the original and shuffled series. The upper (bottom) dashed line represents the maximum (minimum) value of $C_{js}$ as a function of $H_s$ for $d=5$ {and the arrows are indicating the shuffled analysis}.}\label{fig:intgenre}
\end{figure}

Therefore, we have verified that the ordinal pattern distribution that exists among the sound amplitudes values and also among the sound intensity is capable to spread out our database songs though the complexity-entropy causality plane. It is interesting to remark that the embedding dimension employed here ($d=5$) corresponds to approximately $10^{-4}$ seconds. Thus, it is surprising how this very short time dynamics retains so much information about the songs. We also investigated shuffled version of each song series aiming to verify if the localization of the songs in the complexity-entropy causality plane is directly related to the presence of correlations in the music time series. This analysis is shown in Figs.~\ref{fig:ampgenre} and \ref{fig:intgenre} for each song and for all genres. We have obtained $H_s\approx 1$ and $C_{js}\approx 0$ for all shuffled series, confirming that correlations inherently present in the original songs are the main source for the different locations in this plane.

\begin{figure}[!t]
\centering
\includegraphics[scale=0.71]{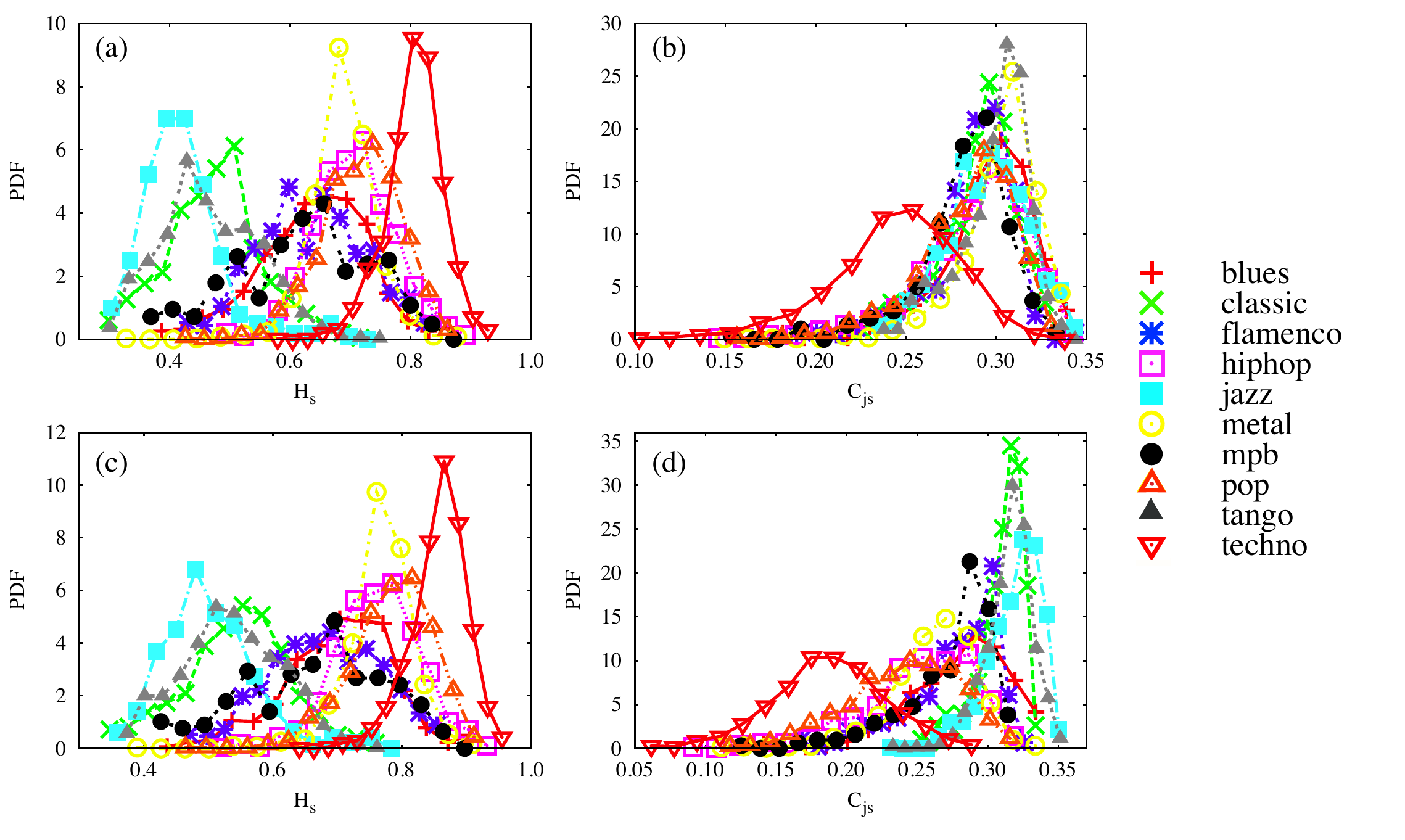}
\caption{(color online) Probability distribution functions (PDF) for the values of (a) $H_s$ and (b) $C_{js}$ when considering the amplitude series grouped by music genre. Figs. (c) and (d) show the same PDFs for the intensity series.}\label{fig:pdfs}
\end{figure}

Although our approach is not focused on determining which music genre is related to a particular given song, this novel physical method may help to understand the complex situation that emerges in the problem of automatic genre classification. For instance, we can take a glance on the fuzzy boundaries existent in the music genre definitions, by evaluating the distribution of $H_s$ and $C_{js}$ values. Figure \ref{fig:pdfs} shows these distributions for both time series employed here. There are several overlapping regions among the distributions of $H_s$ and $C_{js}$ for the different genres. This overlapping
is an illustration on how fuzzy the boundaries between genres and, consequently, the own concept of music genre can be. It is also interesting to observe that some genres have more localized PDFs, for instance, the techno genre is practically bounded to the interval $(0.85,0.95)$ of $H_s$ values for the intensity series while the flamenco or mpb genres have a wider distribution. {To go beyond the previous analysis, we try to quantify the efficiency of permutation indexes $H_s$ and $C_{js}$ in a practical scenery of automatic genre classification. In order to do this, we use an implementation~\cite{SVM1} of a support vector machine (SVM)~\cite{SVM2} where we have considered the values of $H_s$ and $C_{js}$ for the amplitudes and intensity series as features of the SVM. We run the analysis for each genre training the SVM with 90$\%$ of dataset and performing an automatic detection over the remaining 10$\%$. It is a simplified version of the SVM, where the system have to make a binary choice, i.e., to choose between a given genre and all the others. The accuracy rates of automatic detection are shown in Table \ref{tab:SVM}. Note that the accuracy values are around 90$\%$ within this simplified implementation, however we have to remark that in a multiple choice system these values should be much smaller.
On the other hand, this analysis indicates that the entropic indexes employed here may be used in practical situations.
}

\begin{table}
\centering
\caption{Percentage of correct choices of the SVM for each genre}
\label{tab:SVM}
\begin{tabular}{lrclr}
\hline\noalign{\smallskip}
Genre & Accuracy & & Genre & Accuracy\\
\noalign{\smallskip}\hline\noalign{\smallskip}
Blues & 87.87$\%$ & & Metal & 89.89$\%$\\
Classic & 92.03$\%$ & & MPB & 97.15$\%$\\
Flamenco & 95.12$\%$ & & Pop & 88.11$\%$\\
Hiphop & 88.11$\%$ & & Tango & 87.87$\%$\\
Jazz & 91.68$\%$ & & Techno & 87.14$\%$\\
\noalign{\smallskip}\hline
\end{tabular}
\end{table}


\section{Summary and Conclusions}
Summing up, in this work we applied the permutation entropy~\cite{Bandt}, $H_s$, and an intensive statistical complexity measure~\cite{LopezRuiz,Martin,Lamberti}, $C_{js}$, to differentiate songs. Specifically, we analyzed the location 
of the songs in the complexity-entropy causality plane. This permutation information theory approach enabled us to quantitatively classify songs in a kind of complexity hierarchy. 

We believe that the findings presented here may be applied in practical situations as well as in technological applications related to the distinction of songs in massive databases. In this aspect, the Bandt and Pompe approach has some advantageous 
technical features, such as its simplicity, robustness, and principally a very fast numerical evaluation.

\section*{Acknowledgements}
{The authors would like to thank an anonymous reviewer for his very helpful comments. Dr. Osvaldo A. Rosso is also acknowledged for useful discussions and valuable comments.} HVR, RSM and EKL are grateful to CNPq and CAPES (Brazilian agencies) for the financial support. HVR also thanks Angel A. Tateishi for the help with the music database and CAPES for financial support under the process No~5678-11-0. 
LZ was supported by Consejo Nacional de Investigaciones Cient\'ificas y T\'ecnicas (CONICET), Argentina.

\end{document}